\documentclass[12pt]{article}

\usepackage{arxiv}

\usepackage[utf8]{inputenc} % allow utf-8 input
\usepackage[T1]{fontenc}    % use 8-bit T1 fonts
\usepackage{hyperref}       % hyperlinks
\usepackage{url}            % simple URL typesetting
\usepackage{booktabs}       % professional-quality tables
\usepackage{amsfonts}       % blackboard math symbols
\usepackage{nicefrac}       % compact symbols for 1/2, etc.
\usepackage{microtype}      % microtypography
\usepackage{lipsum}
\usepackage{graphicx}
\graphicspath{ {./images/} }
\usepackage{csquotes}

\usepackage{natbib}
\usepackage{tabularx}
\usepackage{rotating}
\usepackage{comment}
\usepackage{setspace}
\usepackage{enumerate}
\title{Cracking the Code: Evaluating Zero-Shot Prompting Methods for Providing Programming Feedback}

\author{Niklas Ippisch, Anna-Carolina Haensch, Jan Simson, Jacob Beck, Markus Herklotz, Malte Schierholz}

\begin{document}

\maketitle

\begin{abstract}
    Despite the growing use of large language models (LLMs) for providing feedback, limited research has explored how to achieve high-quality feedback. This case study introduces an evaluation framework to assess different zero-shot prompt engineering methods. We varied the prompts systematically and analyzed the provided feedback on programming errors in R. The results suggest that prompts suggesting a stepwise procedure increase the precision, while omitting explicit specifications about which provided data to analyze improves error identification.
\end{abstract}

\section{Background}
Large language models (LLMs) like GPT are increasingly being used to support students and developers with programming tasks (Pankiewicz \& Baker, 2023). However, despite their widespread adoption, how to elicit the most helpful feedback from these models is often not addressed in existing research (Dai et al., 2023). To address this gap, we conducted a focused case study, systematically varying and analyzing prompt engineering techniques (Marvin et al., 2024) that could influence the quality of programming-related feedback. 

This evaluation is needed since the results of prompt engineering evaluations are contradictory: some argue that prompt engineering is not task-dependent (Sahoo et al., 2024), while others report big differences between different prompts (Wang et al., 2024; Jacobsen \& Weber, 2024; Mungoli, 2023) or find no or minimal impact on the performance (Tang et al., 2024). Moreover, systematic analyses of the performance of different prompts in differing settings do not exist yet. As a result, it remains uncertain whether the quality of responses resulting from different prompt engineering strategies is context-sensitive, underscoring the need for tests for specific use cases.

In our case, we are interested in how to best elicit feedback on basic R errors typically made by beginners. R, a statistical programming language, is well-known for its steep learning curve at the start: Research suggests an initially high but fast decreasing level of frustration (Baumer et al., 2014) as well as decreasing negative attitudes towards learning R (Tucker et al., 2023) and R outputs (Rode \& Ringel, 2019) during introductory courses. Moreover, it is the first programming experience for many students and therefore much literature exists on how to best teach data analysis and minimize frustration (Tucker et al., 2023; Çetinkaya-Rundel \& Rundel, 2018). Our approach is to equip students with an LLM to provide concise and constructive feedback on their work in R. In that context, the quality of the provided feedback is crucial. To assess the quality of feedback provided by an LLM after varying prompts, we created a systematic evaluation framework. Our contribution is twofold: first, we present our evaluation framework which utilizes the feedback frame of Ryan et al. (2020) and can be applied beyond our study to similar contexts; second, we apply it to a focused use case, analyzing feedback on common beginner errors in R programming. The next section outlines the methodology used for the evaluation, followed by a presentation of the results. The paper concludes with a discussion and closing remarks.

\section{Methods and Data}
 To minimize costs and response times, we limited our evaluation to comparing different zero-shot prompting methods: \textit{Chain of Thought prompting}, \textit{Prompt Chaining}, \textit{Tree of Thought prompting}, and \textit{ReAct prompting} (see the appendix for the exact wording and Marvin et al., 2024 for a summary of prompt engineering methods). In zero-shot \textit{Chain of Thought prompting} a stepwise process is suggested; these steps, in the form of sub-tasks, are made explicit in \textit{Prompt Chaining}. \textit{Tree of Thought prompting} divides the task not in sub-tasks but in different areas where the GPT model operates in one after another and drops the irrelevant. Lastly, \textit{ReAct prompting} enforces reasoning by dividing the task into different thoughts and actions (Yao et al., 2023). We also included a \textit{vanilla model} for comparison.

Two of the authors, who have taught R programming courses for several years, identified five common errors encountered by beginners: (1) false/missing directory, (2) a necessary package is not loaded, (3) prior code is written but not executed, (4) typos in the code, and (5) inconsistent variable naming. We used those as test cases for which the LLM is supposed to provide feedback and suggestions for improvement. In an R environment, we introduced "noise" by adding unrelated code alongside the test cases, deliberately crafted to include specific errors (see the appendix for details\footnote{The introduced noise, the used test codes and prompt wordings can also be found on OSF: osf.io/cgzk9}). After executing the erroneous code, the error message and accompanying environment information were sent to GPT-3.5-turbo through Microsoft Azure alongside the prompt. We checked the response for the evaluation criteria detailed below and cleaned the environment before the next iteration. 

 For the quality of the response, the first and foremost aspect is whether the problem in the script has been detected or not (a). In order to take the quality into account in the analysis, the responses were assessed based on the feedback framework of Ryan et al. (2020). The authors suggest that next to just stating if something is right or wrong, feedback focusing on the response as well as on the underlying concept is also essential for meaningful feedback. Hence, we will also investigate if the concrete error has been described (b) and if it is explained why that error is occurring (c). Additionally for our case of users trouble-shooting problems, the response is supposed to contain a suggestion on how to solve the error (d), be concise (e), and not confuse beginners with suggestions not related to fixing the error (f). In total, that leads to the following six evaluation criteria:

\begin{enumerate}[(a)]
     \item Has the problem been located ('Where is the problem?')?
     \item Has the concrete error been described ('What is the problem?')?
     \item Was the reason for the error explained ('Why is it a problem?')?
     \item Was a suggestion of how to improve given ('How can the problem be solved?')?
     \item Is the feedback concise (below 200 tokens)?
     \item Does the feedback include \textit{only} relevant suggestions and no irrelevant ones (i.e., suggestions which are not directly related to the problematic code)?
 \end{enumerate}

If it was possible to answer the question with 'Yes', one point per iteration was given. If not, then zero points were awarded for that response. Thus, with ten iterations per problem 60 total points were possible.

\section{Results}
\begin{figure}[h!]
    \centering
    \includegraphics[width=0.7\linewidth]{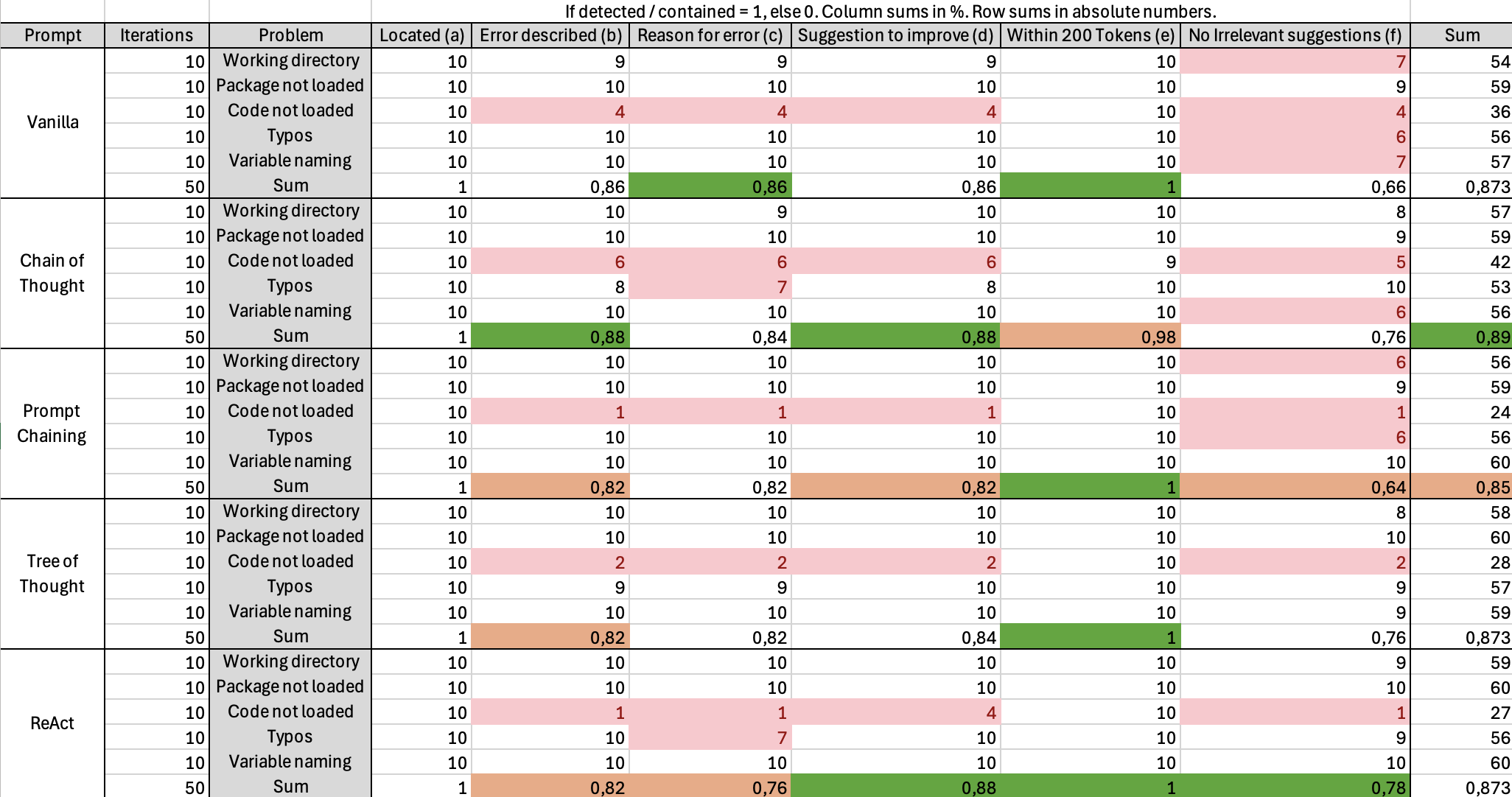}
    \caption{Results of the evaluation}
    \label{fig:results_evaluation}
\end{figure}

Figure \ref{fig:results_evaluation} shows the evaluation results. We iterated ten times per prompt and problem. The cells show the absolute frequencies of how often we were able to answer the above questions with 'Yes'. All cells that are shaded pink indicate absolute frequencies below 8. The orange cells indicate the worst performance per criteria and the green cells the best, respectively. For every type of prompt, the bottom line shows the column percentage and the last column depicts the line sums. For assessing the overall performance (bottom right cell), the total number of 'Yes' was divided by the total number of possible points, i.e., by 300. 

Overall, all prompts performed pretty well, with overall ratings above 0.85, but there are subtle differences between the prompts. All responses detected the error, and all but one (\textit{Chain of Thought}) were within 200 tokens. Performance was consistent across prompts for errors such as false working directories, unloaded packages, and inconsistent variable naming. 

Regarding typos and not loaded code, however, the performance varies more. \textit{Chain of Thought prompting} and \textit{ReAct} had bigger problems in providing a correct reason for errors resulting from typos: they assumed that the code \texttt{summary[data\$hp]} is supposed to just access the column hp, i.e., they prioritized the brackets over the \texttt{summary}-function. Whereas \textit{Chain of Thought prompting} performed best among all prompts regarding not loaded code, the same is not true for \textit{ReAct}. All prompts struggled to identify cases where the required data column was missing, even though the line of code was already written. Since they detected the missing column, \textit{Chain of Thought prompting} and the \textit{vanilla prompt} appear to take the information provided more into account.

The last bigger difference is visible regarding irrelevant suggestions. \textit{Chain of Thought prompting, Tree of Thought prompting,} and \textit{ReAct} were more able to avoid suggestions in their response that are not directly related to the error message, e.g. the recommendation to check all packages attached in a typo-error.

To summarize, it seems like there is a trade-off between a higher precision (i.e., no irrelevant suggestions) and a better detection of the underlying error (i.e., higher scores in terms of error identification and reason for error). Regarding the \textbf{precision}, prompts that enforce a stepwise process (\textit{Chain of Thought, Tree of Thought,} and \textit{ReAct}) seem to perform better. Prompts, in which the information to be analyzed (e.g., script, data, etc.) are not specifically mentioned (\textit{Vanilla} and \textit{Chain of Thought}), seem to perform better regarding the \textbf{error identification and reason}. Two exceptions are visible: \textit{Prompt Chaining}, where a stepwise process is enforced, and the data to be analyzed is mentioned, performs less both in terms of precision and error identification. On the other hand, the \textit{Chain of Thought prompting} excelled in precision despite not explicitly specifying information to analyze.

\section{Concluding remarks}
Due to the uncertain context sensitivity of prompt engineering strategies, it was necessary to evaluate them for our specific use case. To provide high-quality feedback on errors in R codes, we analyzed four different zero-shot prompting strategies as well as a vanilla prompt. The results show in our case that all prompts performed well and were pretty similar. However, there exist differences regarding how much different prompts take into account the information provided along them. The data suggests that the precision increases when a stepwise procedure is enforced, and the error identification improves when the data to be analyzed is not explicitly mentioned. The latter seems contra-intuitive. Apparently, the LLM checks more reliably on its own whether the column exists in the data provided than when the data is mentioned along all the other information to check (e.g., script, packages, directory, ...). Thus, it seems mentioning all possible information sources rather leads to a confusion of the LLM. 

Our proposed evaluation framework provides a structured and replicable method to assess prompt engineering strategies for programming assistance. While our study focuses on errors commonly made by beginners in R programming, the framework is designed to be applicable across various use cases. Researchers and practitioners can adopt and adapt this pipeline to evaluate LLM feedback quality for other programming languages or advanced coding problems.  We encourage others to build on our approach, refining and extending it to further enhance the effectiveness of LLMs in educational and professional settings.

\section*{References}
Baumer, B., Çetinkaya-Rundel, M., Bray, A., Loi, L. \& Horton, N. J. (2014). R Markdown: Integrating A Reproducible Analysis Tool into Introductory Statistics. https://arxiv.org/pdf/1402.1894 \\
Çetinkaya-Rundel, M. \& Rundel, C. (2018). Infrastructure and Tools for Teaching Computing Throughout the Statistical Curriculum. \textit{The American Statistician, 72}(1), 58-65.  https://doi.org/10.1080/00031305.2017.1397549 \\
Dai, W., Lin, J., Jin, F., Li, T., Tsai, Y. S., Gasevic, D. \& Chen, G. (2023). Can Large Language Models Provide Feedback to Students? A Case Study on ChatGPT. \textit{2023 IEEE International Conference on Advanced Learning Technologies (ICALT)}. https://doi.org/10.1109/ICALT58122.2023.00100 \\
Jacobsen, L. J. \& Weber, K. E. (2024). The Promises and Pitfalls of LLMs as Feedback Providers: A Study of Prompt Engineering and the Quality of AI-Driven Feedback. Preprint. https://doi.org/10.31219/osf.io/cr257 \\
Marvin, G., Hellen, N., Jjingo, D. \& Nakatumba-Nabende, J. (2024). Prompt Engineering in Large Language Models. In: Jacob, I. J., Piramuthu, S. \& Falkowski-Gilski, P. (Eds.). Data Intelligence and Cognitive Informatics. Proceedings of ICDICI 2023, 387-402, Springer. \\
Mungoli, N. (2023). Exploring the Synergy of Prompt Engineering and Reinforcement Learning for Enhanced Control and Responsiveness in Chat GPT. \textit{Journal of Electrical Electronics Engineering, 2}(3), 201-205. https://doi.org/10.33140/JEEE.02.03.02 \\
Pankiewicz, M. \& Baker, R. S. (2023). Large Language Models (GPT) for automating feedback on programming assignments. \textit{Proceedings of the 31st International Conference on Computers in Education. Asia-Pacific Society for Computers in Education}. https://arxiv.org/pdf/2307.00150 \\
Rode, J. B. \& Ringel, M. M. (2019). Statistical Software Output in the Classroom: A Comparison of R and SPSS. \textit{Teaching of Psychology, 46}(4), 319-327. https://doi.org/10.1177/0098628319872605 \\
Ryan, A., Judd, T., Swanson, D., Larsen, D. P., Elliott, S., Tzanetos, K. \& Kulasegaram, K. (2020). Beyond right or wrong: More effective feedback for formative multiple-choice tests. \textit{Perspectives on Medical Education, 9}(5), 307-313. https://doi.org/10.1007/s40037-020-00606-z \\
Sahoo, P., Singh, A. K., Saha, S., Jain, V., Mondal, S. \& Chadha, A. (2024). A Systematic Survey of Prompt Engineering in Large Language Models: Techniques and Applications. arXiv:2402.07927v1 \\
Tang, Y., Xia, Z., Li, X., Zhang, Q., Chan, E. W. \& Wong, I. CK. (2024). Large Language Model in Medical Information extraction from Titles and Abstracts with Prompt Engineering Strategies: A Comparative Study of GPT-3.5 and GPT-4. Preprint. https://doi.org/10.1101/2024.03.20.24304572 \\
Tucker, M. C., Shaw, S. T., Son, J. Y. \& Stigler, J. W. (2023). Teaching Statistics and Data Analysis with R. \textit{Journal of Statistics and Data Science Education, 31}(1), 18-32. https://doi.org/10.1080/26939169.2022.2089410 \\
Wang, L., Chen, X., Deng, X. W., Wen, H., You, M. Liu, W. \& Li, J. (2024). Prompt engineering in consistency and reliability with the evidence-based guidline for LLMs. \textit{npj Digital Medicine, 7}. https://doi.org/10.1038/s41746-024-01029-4 \\
Yao, S., Zhao, J., Yu, D., Du, N., Shafran, I., Narasimhan, K. \& Cao, Y. (2023). ReAct: Synergizing Reasoning and Acting in Language Models. 	arXiv:2210.03629

\newpage
\section*{Appendix}
\textbf{Test codes:} \\
(1) False working direcotry \\
Code: \hspace{2.05cm} \texttt{read.csv("data/testfile")} \\
Problem: \hspace{1.6cm} no folder data and no file ‘testfile’ \\
Error message: \hspace{0.8cm} \texttt{Error in file(file, "rt) : cannot open the connection} \\

(2) Package not loaded \\
Code: \hspace{2.05cm} \texttt{plot = ggplot(data = mtcars, mapping = aes(x = hp, y = cyl)) +} \\
\hspace*{3cm} \texttt{geom\_point()} \\
Problem: \hspace{1.6cm} package \texttt{ggplot2} not loaded \\
Error message: \hspace{0.8cm} \texttt{Error in ggplot(data = mtcars, mapping = aes(x = hp, y = cyl)) : could }  \\
\hspace*{3cm} \texttt{not find function "ggplot"}

(3) Code not loaded \\
Code: \hspace{2.05cm} \texttt{data = mtcars} \\
\hspace*{3cm} \texttt{data\$weight\_kg = data\$wt*0.454} \\
\hspace*{3cm} \texttt{data\$weight\_kgsq = data\$weight\_kg \^2} \\
Problem: \hspace{1.6cm} the middle-line is not executed, hence the column \texttt{weight\_kg} does not exist \\
Error message: \hspace{0.8cm} \texttt{Error in '\$<-.data.frame'('*tmp*, weight\_kgsq, value = numeric(0)) : }   \\
\hspace*{3cm} \texttt{replacement has 0 rows, data has 32}

(4) Typos \\
Code: \hspace{2.05cm} \texttt{data = mtcars} \\
\hspace*{3cm} \texttt{summary[data\$hp]} \\
Problem: \hspace{1.6cm} brackets instead of parentheses \\
Error message: \hspace{0.8cm} \texttt{Error in summary[data\$hp] : object of type 'closure' is not subsettable}  \\

(5) Variable naming \\
Code: \hspace{2.05cm} \texttt{var1 = 5} \\
\hspace*{3cm} \texttt{var2 = 3} \\
\hspace*{3cm} \texttt{var3 = var\_1 * var2} \\
Problem: \hspace{1.6cm} in the third line, \texttt{var\_1} instead of \texttt{var1} is used \\
Error message: \hspace{0.8cm} \texttt{Error: object 'var\_1' not found} \\

\newpage
\textbf{Prompts}

System prompt (kept constant for all of the prompts below): \\
"You are helping students in an R programming course for beginners and give feedback on why it is wrong, how to correct it and how to improve in the future."

(1) Vanilla prompt \\
"Identify the errors (there might be multiple) and give me feedback on how to correct the issue in maximum three sentences."

(2) Chain of Thought \\
"Analyze the information provided step by step and afterwards give feedback on how to correct the issue in maximum three sentences."

(3) Prompt Chaining \\
"In the first step, analyze the script provided, loaded data and their structure, existing variables and functions and the packages loaded. In the second step, use the error message and the relevant parts of your analysis of the information and give feedback on how to correct the issue in maximum three sentences."

(4) Tree of Thought \\
"Assume, the error is in one of the following areas: written script and code, loaded data, loaded variables and functions, packages. Stepwise, check whether there is an error in each of the areas connected to the error message provided. If not, drop that area. When just one area is left, provide feedback on how to correct the issue in maximum three sentences."

(5) ReAct \\
"Thought 1 I first have to analyze the information provided and screen briefly the script, the loaded data including the structure, the loaded packages, and the variables and functions created. \\
Action 1 Analyze information \\
Thought 2 An error message is mentioned. I should check the error message and identify, based on the prior analysis, the relevant parts of the information provided which are linked to the error message. \\
Action 2 Check error message \\
Thought 3 After the analysis of the information and error message, I am now able to concisely explain why the error message occurred and how to correct it. \\
Action 3 End [Explain] \\
Question Give feedback on how to correct the issue in maximum three sentences."

\end{document}